\documentclass[11pt]{article} 
\usepackage{graphicx,graphics,floatflt,amssymb,epsf,rotate} 
\usepackage{color}

\def\H{\hat{H}}
\def\L{\hat{L}}
\def\Q{\hat{Q}}
\def\U{\hat{U}}
\def\D{\hat{D}}
\def\E{\hat{E}}

\begin{document} 
\begin{flushright} 
SINP/TNP/2008/15,  
LPT-ORSAY 08-67 
\end{flushright} 
 
\vskip 30pt 
 
\begin{center}
  {\Large \bf Electroweak Symmetry Breaking and BSM Physics (A
    Review)\footnote{ Based on plenary talks at the International Conferences:
      WIN07, Kolkata, Jan'07, and WHEPP-10, Chennai, Jan'08. To appear
      in the WHEPP-10 proceedings (a special issue of PRAMANA).} } \\
  \vspace*{1cm} \renewcommand{\thefootnote}{\fnsymbol{footnote}}
  {\bf Gautam Bhattacharyya} \\
  \vspace{10pt} {\small {\em Saha Institute of Nuclear Physics, 1/AF Bidhan
      Nagar, Kolkata 700064, India}}
    
\normalsize 
\end{center} 

\begin{abstract}  
  
  In this talk, I shall first discuss the standard model Higgs mechanism and
  then highlight some of its deficiencies making a case for the need to go
  beyond the standard model (BSM). The BSM tour will be guided by symmetry
  arguments. I shall pick up four specific BSM scenarios, namely,
  supersymmetry, Little Higgs, Gauge-Higgs unification, and the Higgsless
  approach. The discussion will be confined mainly on their electroweak
  symmetry breaking aspects.

\vskip 5pt \noindent 
\texttt{PACS Nos:~ 12.60.Jv, 11.10.Kk } \\ 
\texttt{Key Words:~~Higgs, Supersymmetry, Extra Dimension}
\end{abstract}

\renewcommand{\thesection}{\Roman{section}} 
\setcounter{footnote}{0} 
\renewcommand{\thefootnote}{\arabic{footnote}}

\section{Introduction}
The understanding of electroweak symmetry breaking (EWSB) would be right at
the top of the agenda when we would start analysing the LHC data
\cite{rev_ewsb,Kaul:2008cv,Cheng:2007bu,Rattazzi:2005di}. Both the CMS and the
ATLAS detectors are poised to resolve this issue. The question is the
following: whether the Higgs mechanism as depicted in the standard model (SM)
is a complete description of EWSB consistent with all experimental data, or
there is a more fundamental underlying dynamics that mimics a Higgs-like
picture at the electroweak scale.  On theoretical grounds, the latter seems to
be the case. Then, in what form would that new physics beyond the standard
model (BSM) manifest in the LHC data? We need to crack several codes before we
can possibly unravel the most {\em expensive} secret challenging our
imagination!

The SM reigns supreme at the electroweak scale and electroweak precision tests
(EWPT), primarily at LEP, have put a lot of restraints on how a BSM scenario
should be perceived. Non-abelian gauge theory has been established to a very
good accuracy: (i) the $ZWW$ and $\gamma WW$ vertices have been measured to a
per cent accuracy at LEP-2 implying that the SU(2) $\times$ U(1) gauge theory
is unbroken at the vertices, (ii) accurate measurements of the $Z$ and $W$
masses have indicated that gauge symmetry is broken in masses, and the
longitudinal polarizations of those gauge bosons, which are absent in the
unbroken phase of the symmetry, should find their ancestry in the dynamics
that portrays a Higgs-like picture. The $\rho$-parameter has been measured to
a very good accuracy as being very close to unity - a feature that attests the
doublet structure of the Higgs assumed in the SM.  Any viable BSM scenario
should be in accord to the above properties. In what follows, we first briefly
review the SM Higgs mechanism and then take on supersymmetry, little Higgs,
gauge-Higgs unification and the Higgsless models, keeping our discussions
confined only to their EWSB aspects.

\section{The SM Higgs mechanism}
There is a complex scalar doublet $\Phi \equiv \left(\phi_1 + i\phi_2,
\phi_3 + i\phi_4 \right)^T$, and the potential is  
\begin{equation} 
\label{smpot}
V(\Phi) =  - \mu^2 \Phi^\dagger \Phi + \lambda (\Phi^\dagger \Phi)^2 \, .
\end{equation}
Spontaneous symmetry breaking (SSB) requires $\mu^2 > 0$, and the stability of
the potential (that it is bounded from below) demands that $\lambda > 0$.
After SSB, an order parameter, called `vacuum expectation value (vev)', is
generated: $v = \sqrt{\mu^2/\lambda}$.  The charged and neutral force
particles, namely, the $W^\pm$ and $Z$ bosons, `swallow' the $(\phi_1,\phi_2)$
and $\phi_4$ components of $\Phi$ to constitute their longitudinal
polarizations, which yield $M_W = gv/2$ and $M_Z = (\sqrt{g^2+g^{\prime
    2}})v/2$, where $g$ and $g'$ are SU(2) and U(1) gauge couplings,
respectively. The fermion masses are also controlled by $v$ and are given by
$m_f = h_f v/\sqrt{2}$, where $h_f$ is the Yukawa coupling of the fermion
$f$. The Higgs boson ($h$) arises from the quantum fluctuation of $\phi_3$
around the vev ($\phi_3 = (v+h)/\sqrt{2}$), and $m_h = \sqrt{2\lambda} v$. The
latest global electroweak fit gives $M_Z = 91.1875 \pm 0.0021$ GeV and $M_W =
80.398 \pm 0.025$ GeV.

\subsection{Constraints on the Higgs mass}
\subsubsection{Electroweak fit}
The Higgs mass enters electroweak fit through the $\Delta \rho$ and $S$
parameters. At the tree level, $\rho^{\rm SM}=1$. The quantum 
corrections show a logarithmic sensitivity to the Higgs mass: 
\begin{eqnarray}
\label{st}
\Delta \rho  \simeq  \frac{3G_F}{8\pi^2\sqrt{2}} 
\left[m_t^2 - (M_Z^2 - M_W^2) \ln(\frac{m_h^2}{M_Z^2}) \right],~~  
 S \simeq \frac{1}{6\pi} \ln\left(\frac{m_h}{M_Z} \right) .
\end{eqnarray}
There is a strong quadratic dependence on the top mass. At present, the CDF
and D0 combined estimate is $m_t = 172.6 \pm 1.4$ GeV. This translates into an
upper limit on the Higgs mass: $m_h < 186$ GeV at 95\% CL (imposing the direct
search limit $m_h > 114.4$ GeV in the fit, from non-observation of Higgs at
LEP-2 in the Bjorken process $e^+e^- \to Zh$) \cite{lepewwg}. 
Figs.~1a and 1b capture the details.
\begin{figure*}
\begin{minipage}[t]{0.42\textwidth}
\epsfxsize=6cm
\centering{\epsfbox
{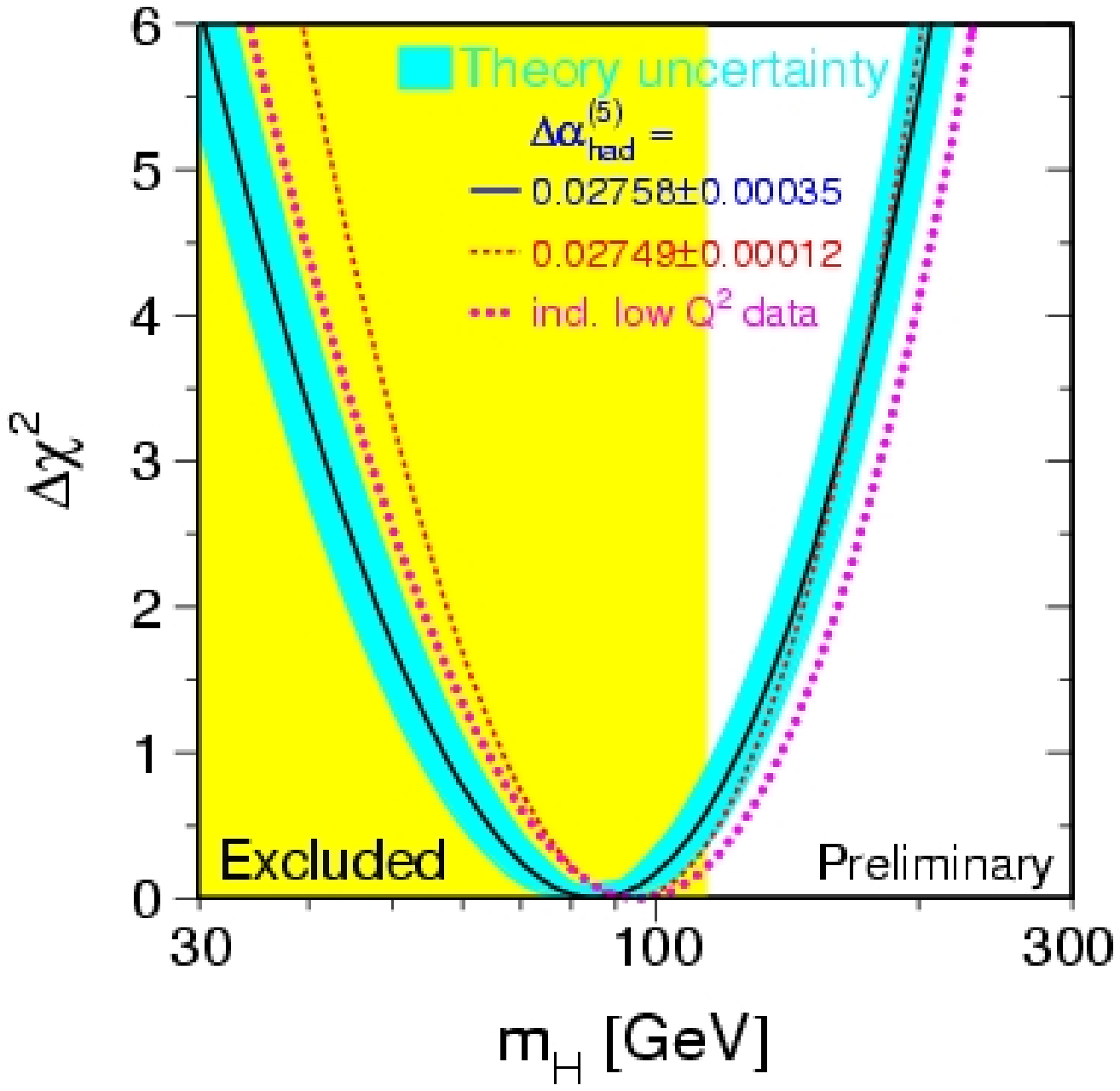}}
\end{minipage}
\hfill
\begin{minipage}[t]{0.42\textwidth}
\epsfxsize=6cm
\centering{\epsfbox
{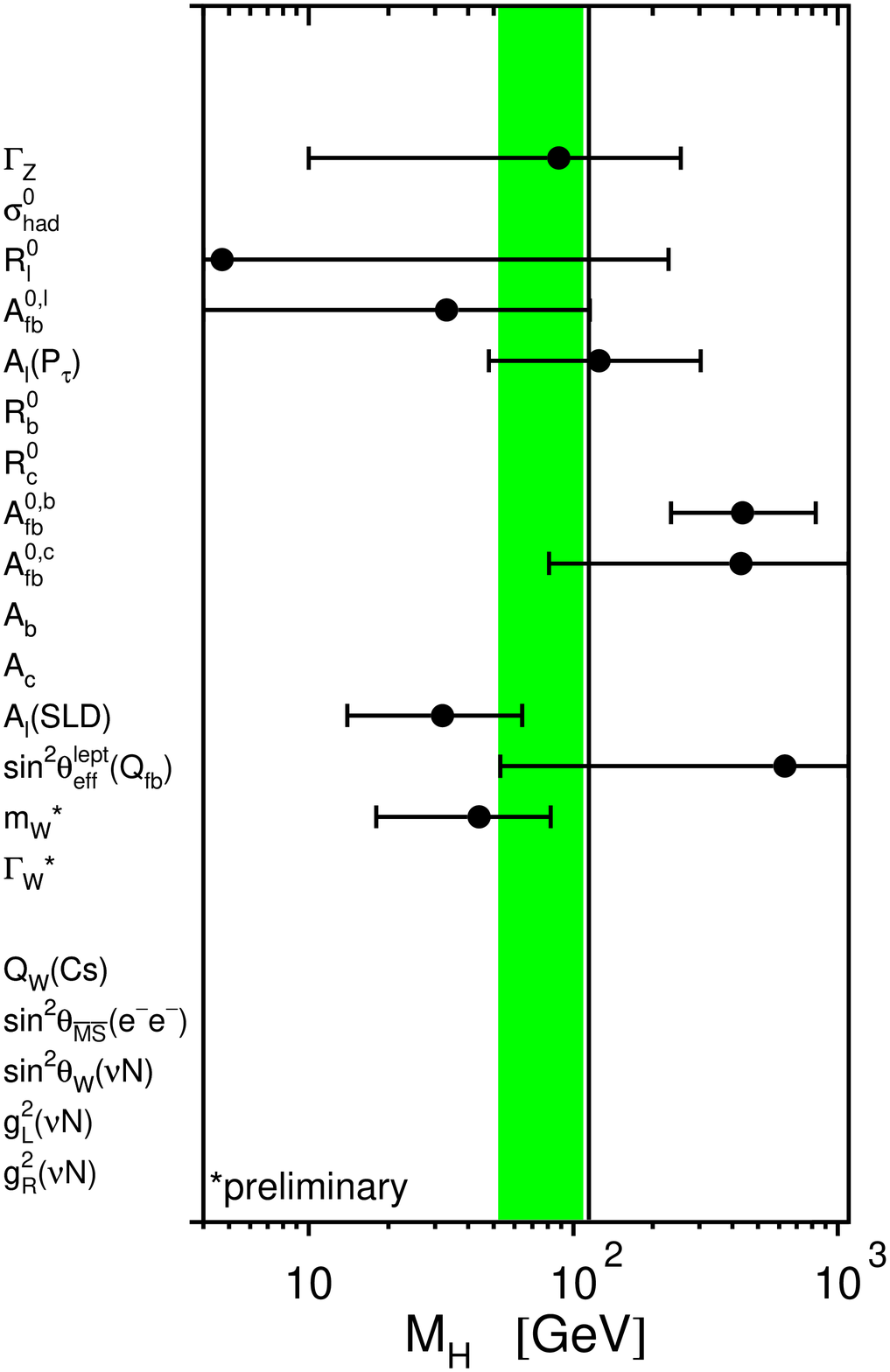}}
\end{minipage}
\vspace{5pt}
\caption[]{\small{(a) Left panel: The blue-band plot showing the Higgs mass
    limits \cite{lepewwg}. (b) The limits on the Higgs mass from different
    measurements. The central band corresponds to the `average'
    \cite{lepewwg}.}}
\end{figure*}

\subsubsection{Theoretical limits} 
{\em Unitarity} \cite{Lee:1977eg} places an upper bound on $m_h$ beyond which
the theory becomes non-perturbative. Here, we shall call it a `tree level
unitarity' as we would require that the tree level contribution of the first
partial wave in the expansion of different scattering amplitudes does not
saturate unitarity (in other words, some probability should not exceed
unity). The scattering amplitudes involving gauge bosons and Higgs can be
decomposed into partial waves (using `equivalence theorem') as
\begin{equation}
a_J (s) = \frac{1}{32\pi} \int d(\cos\theta) P_J (\cos\theta) M(s,\theta) ,
\end{equation}
where $a_J$ is the $J$th partial wave, $P_J$ is the $J$th Legendre polynomial
and $M(s,\theta)$ is the scattering matrix element. The {\em most} divergent
scattering amplitude arises from $2 W_L^+ W_L^- + Z_L Z_L$ channel, leading to
$a_{J=0} = -5 m_h^2/64 \pi v^2$. Satisfying the unitarity constraint,
i.e. $|{\rm Re}~ a_J| \leq 0.5$, yields $m_h < 780$ GeV.

Besides, there are theoretical upper and lower limits on the Higgs mass
arising from the twin requirements \cite{Altarelli:1994rb,Kolda:2000wi}: (i)
the running quartic coupling $\lambda(\mu)$ should not hit the Landau pole
throughout the history of renormalization group (RG) evolution from the
electroweak scale $v$ to some cutoff $\Lambda$, and (ii) $\lambda (\mu)$
should always stay positive, so that the scalar potential remains bounded from
below.  The bounds follow from the following RG evolution of the quartic
coupling, given by ($t=\ln(\mu/v)$):
\begin{equation}
d\lambda/dt = (4\pi^2)^{-1}  
3\left[\lambda^2 + \lambda h_t^2 -  h_t^4 - ...
\right] \, . 
\end{equation}
Recall, $m_h = \sqrt{\lambda(v)} v$, and that is how the Higgs mass enters
into the game. The {\em triviality} argument of staying within the
perturbative limit by maintaining $\lambda^{-1} (\mu) > 0$ leads to an upper
limit $m_h < 170$ GeV for $\Lambda = 10^{16}$ GeV. The {\em vacuum stability}
argument, that $\lambda (\mu) > 0$ bounds the potential from below, sets a
lower limit $m_h > 130$ GeV for $\Lambda = 10^{16}$ GeV. The limits for other
choices of the cutoff can be read off from Figs.~2a and 2b.
\begin{figure*}
\begin{minipage}[t]{0.42\textwidth}
\epsfxsize=6cm
\centering{\epsfbox
{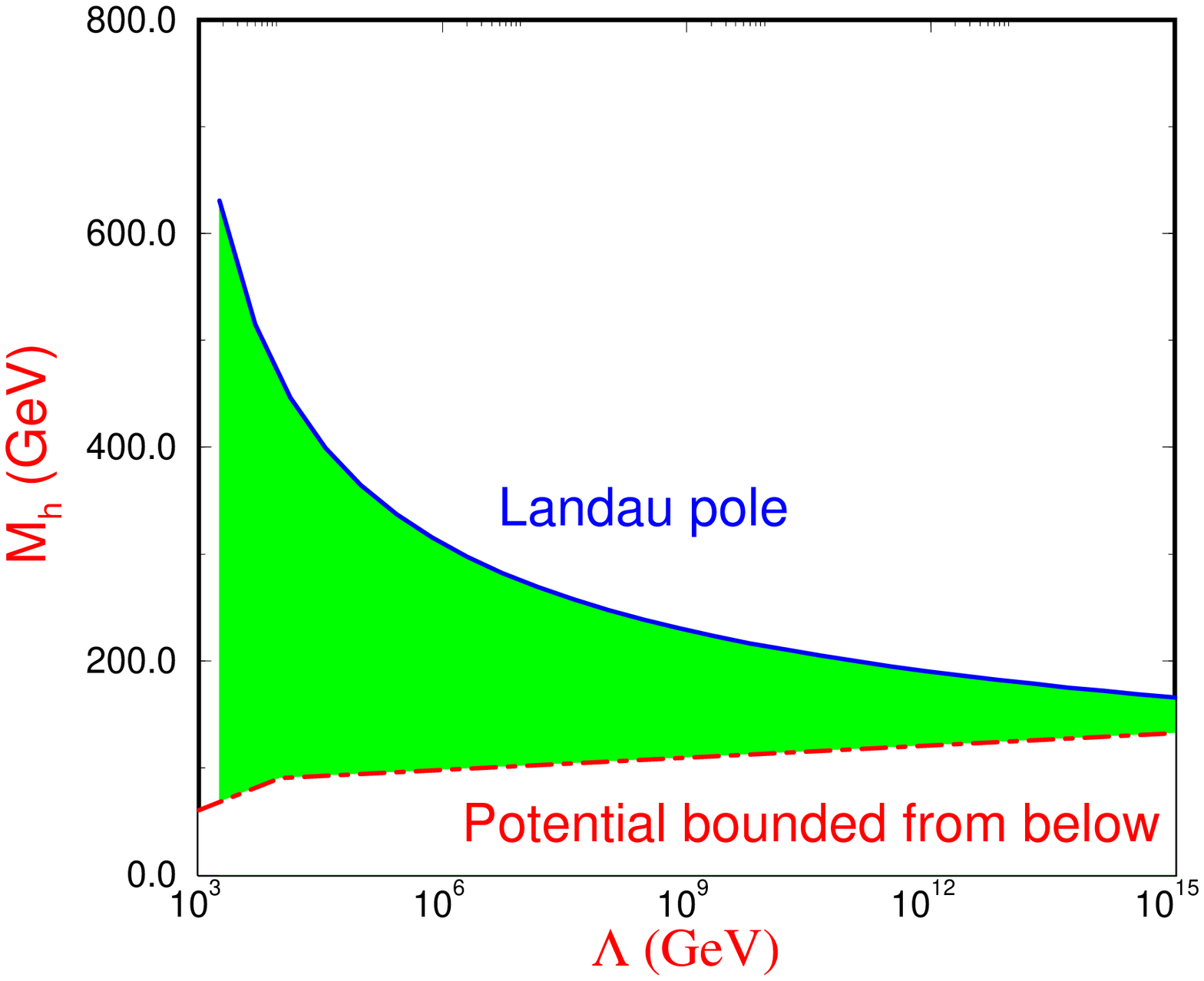}}
\end{minipage}
\hfill
\begin{minipage}[t]{0.42\textwidth}
\epsfxsize=6cm
\centering{\epsfbox
{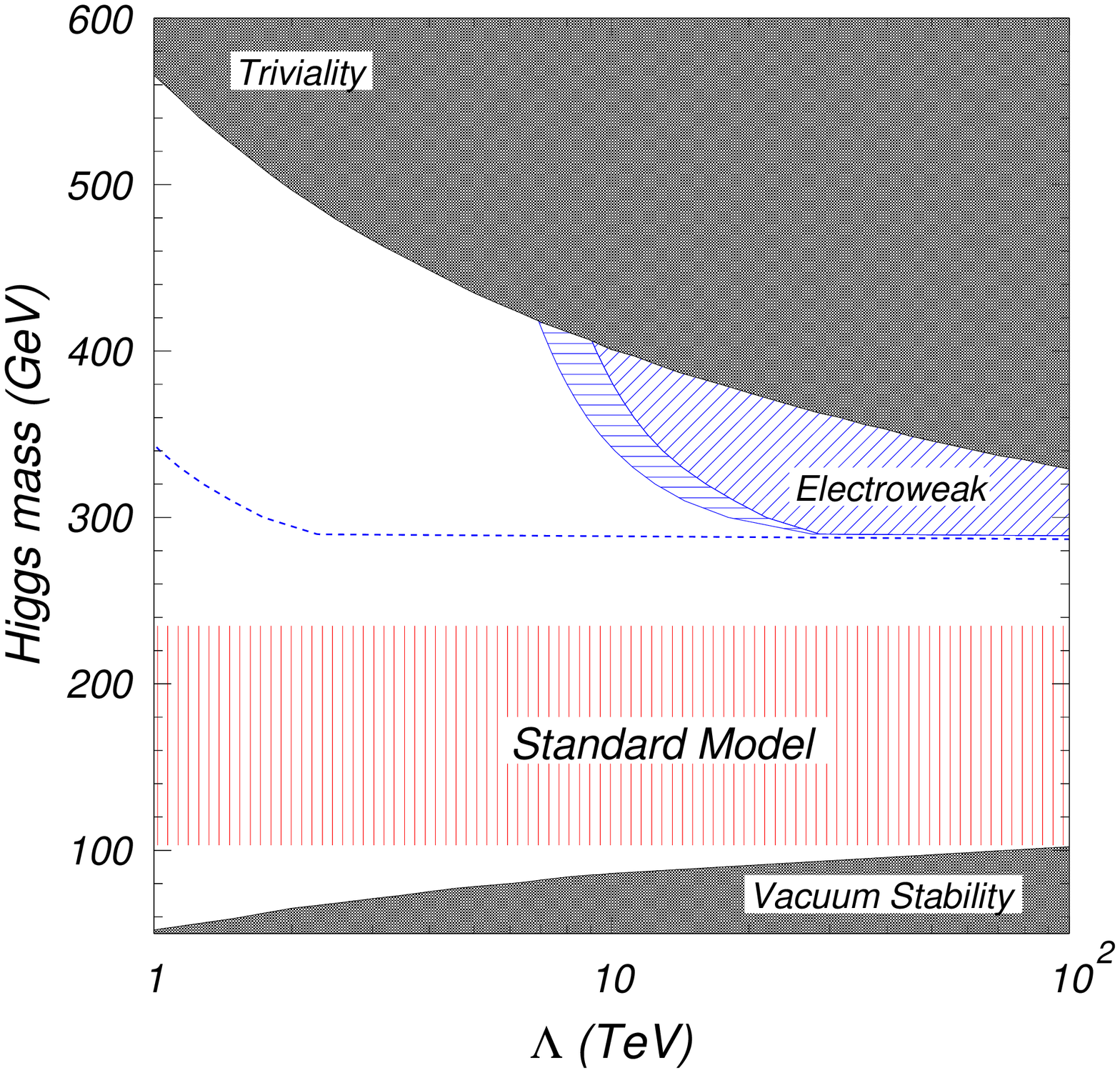}}
\end{minipage}
\vspace{5pt}
\caption[]{\small{(a) Left panel: The triviality and vacuum stability limits
    \cite{Altarelli:1994rb}. (b) Right panel: The region upto $\Lambda = 100$
    TeV is zoomed.  The hatched region `Electroweak' is ruled out by precision
    data. Only the `Standard Model' region is allowed at 95\%
    CL \cite{Kolda:2000wi}.}}
\end{figure*}

\subsection{Will the discovery of Higgs mark the end of the story?}
Once we discover the Higgs, the SM spectrum is completed. But, will a
completion be achieved in terms of our understanding of the universe through
elementary particle interactions? We should remember that there are phenomena
which cannot be explained by the SM: notably, the neutrino mass, the dark
matter and the acceleration of the universe. Moreover, there is a conceptual
loop-hole in the description of the scalar sector: the SM suffers from the
{\em gauge hierarchy problem}. The quantum correction to the Higgs mass is
quadratically sensitive to the cutoff,
\begin{equation}
\label{quadcorr}
\Delta m_h^2 (f)  =  - \frac{h_f^2}{16\pi^2} 2 \Lambda^2 ~~~~~,~~~~~ 
\Delta m_h^2 (S)  =   \frac{\lambda}{16\pi^2}  \Lambda^2 \, ,
\end{equation}
where $f$ and $S$ in brackets stand for fermionic and scalar loops. The cutoff
dependence arises because no symmetry protects the Higgs mass. Recall that in
QED the electron mass is protected by chiral symmetry, $\Delta m_e = m_e
\frac{\alpha}{4\pi} \ln(\Lambda)$, so that $m_e \to 0$ gives an enhanced
symmetry. In the electroweak theory after the SSB, as we can see from
Eq.~(\ref{quadcorr}), there is no such enhanced symmetry when $m_h \to
0$. More precisely, the vev $v$ is not protected from {\em large} quantum
corrections. Thus, while on one hand we demand the Higgs to weigh around a few
hundred GeV, on the other hand the quantum correction pushes it up to the
cutoff (e.g. the GUT scale). Even if we absorb the one-loop $\Lambda^2$ terms
by a redifinition of $\mu^2$ or tuning $\lambda = 2 h_f^2$, when we go to
two-loop the $\Lambda$-dependence again shows up with different coefficients,
and thus we need to tune the parameters again. We have to repeat it
order-by-order in perturbation theory, which makes the theory meaningless.
This is what constitutes the gauge hierarchy problem. Since $v$ is not stable,
not only the Higgs mass, the masses of the gauge bosons and fermions are not
stable either.

Another bothering issue is the {\em negative} sign put {\em by hand} in front
of $\mu^2$ in Eq.~(\ref{smpot}) to make SSB happen. We must have a dynamical
understanding of this {\em ad hoc} `minus sign'.

Now we shall take a few examples of BSM scenarios and observe how in each case
some underlying symmetry regulates the Higgs mass.

\section{Supersymmetry} 
\subsection{Basics}
Supersymmetry is a new space-time symmetry interchanging bosons and fermions,
relating states of different spin \cite{susy-books,reviews}.  The Poincare
group is extended by adding two anticommuting generators $Q$ and $\bar{Q}$, to
the existing $p$ (linear momentum), $J$ (angular momentum) and $K$ (boost),
such that $\{Q,\bar{Q}\} \sim p$. Since the new symmetry generators are
spinors, not scalars, supersymmetry is not an internal symmetry. Recall, Dirac
postulated a doubling of states by introducing an antiparticle to every
particle in an attempt to reconcile Special Relativity with Quantum Mechanics.
In Stern-Gerlach experiment, an atomic beam in an inhomogeneous magnetic field
splits due to doubling of the number of electron states into spin-up and -down
modes indicating a doubling with respect to angular momentum. So it is no
surprise that $Q$ would cause a further splitting into particle and
superparticle ($f \stackrel{Q}{\rightarrow} f, \widetilde{f}$)
\cite{hall}. Since $Q$ is spinorial, the superpartners differ from their SM
partners in spin. The superpartners of fermions are scalars, called
`sfermions', and those of gauge bosons are fermions, called `gauginos'. Put
together, a particle and its superpartner form a supermultiplet. The two
irreducible supermultiplets which are used to construct the supersymmetric
standard model are the `chiral' and the `vector' supermultiplets. The chiral
supermultiplet contains a scalar (e.g. selectron) and a 2-component Majorana
fermion (e.g. left-chiral electron). The vector supermultiplet contains a
gauge field (e.g. photon) and a 2-component Majorana fermion (e.g. photino).
Two points are worth noting: (i) there is an equal number of bosonic and
fermionic degrees of freedom in a supermultiplet; (ii) since $p^2$ commutes
with $Q$, the bosons and fermions in a supermultiplet are mass degenerate.

\subsection{Motivation}

\subsubsection{Supersymmetry solves the gauge hierarchy problem }

An attempt to solve the `gauge hierarchy problem', i.e., why $M_{\rm Pl} \gg
M_W$, or equivalently, $G_N \ll G_F$, is the main motivation behind the
introduction of supersymmetry \cite{intro-susy}.  We recall from the previous
section that quantum corrections to the Higgs mass from a bosonic loop and a
fermionic loop have {\em opposite} signs.  So if the couplings are identical
and the boson is mass degenerate with the fermion, the net contribution would
vanish!  What can be a better candidate than supersymmetry to do this job? For
every particle supersymmetry provides a mass degenerate partner differing by
spin $1\over 2$.  However, the cancellation is not exact because in real world
supersymmetry is badly broken. But if the breaking occurs through `soft'
terms, i.e. in masses and not in couplings, the quadratic divergence still
cancels. The residual divergence is mild, only logarithmically sensitive to
the supersymmetry breaking scale.

\subsubsection{Supersymmetry leads to unification of gauge couplings}

This was a bonus \cite{uni}. Supersymmetry was introduced not with this in
mind!  In the SM, when the gauge couplings are extrapolated to high scale,
with LEP measurements as input values, they do not meet at
a single point. In supersymmetry, they do, at a scale $M_{\rm GUT} \sim 2
\times 10^{16}$ GeV, provided the superparticles weigh around 1 TeV (see
Fig.~3a).

\subsubsection {Supersymmetry triggers EWSB}

The mass-square of one of the Higgs, $m_{H_u}^2$, starting from a positive
value in the ultraviolet becomes negative in the infrared triggering EWSB. As
we remarked in the previous section that in the SM the negative sign in front
of the scalar mass-square in the potential is completely {\em ad hoc} and put
in by hand to ensure EWSB. In supersymmetry it is the heavy top quark that
radiatively induces the sign flip (see Fig.~3b).
\begin{figure*}
\begin{minipage}[t]{0.42\textwidth}
\epsfxsize=6cm
\centering{\epsfbox
{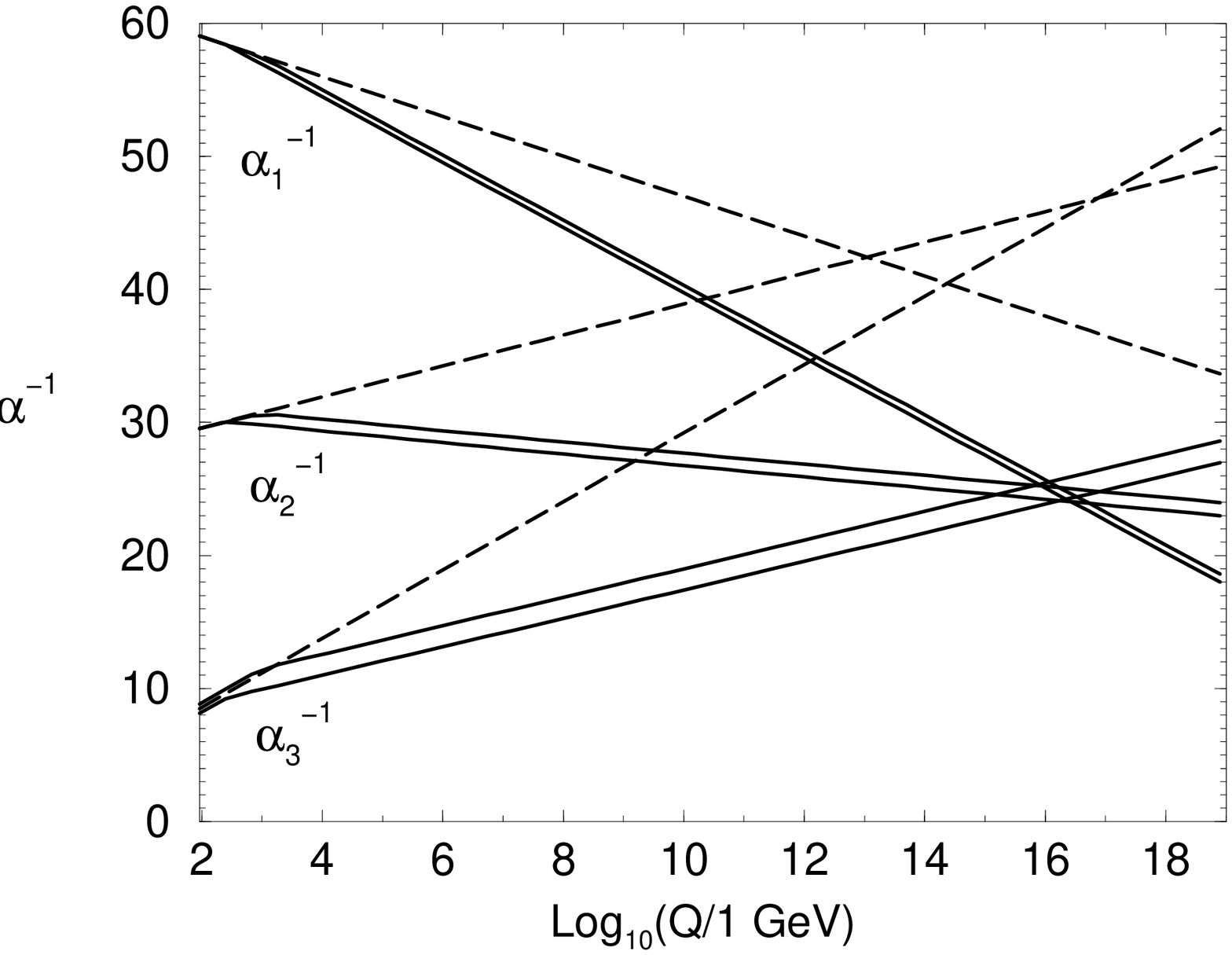}}
\end{minipage}
\hfill
\begin{minipage}[t]{0.42\textwidth}
\epsfxsize=6cm
\centering{\epsfbox
{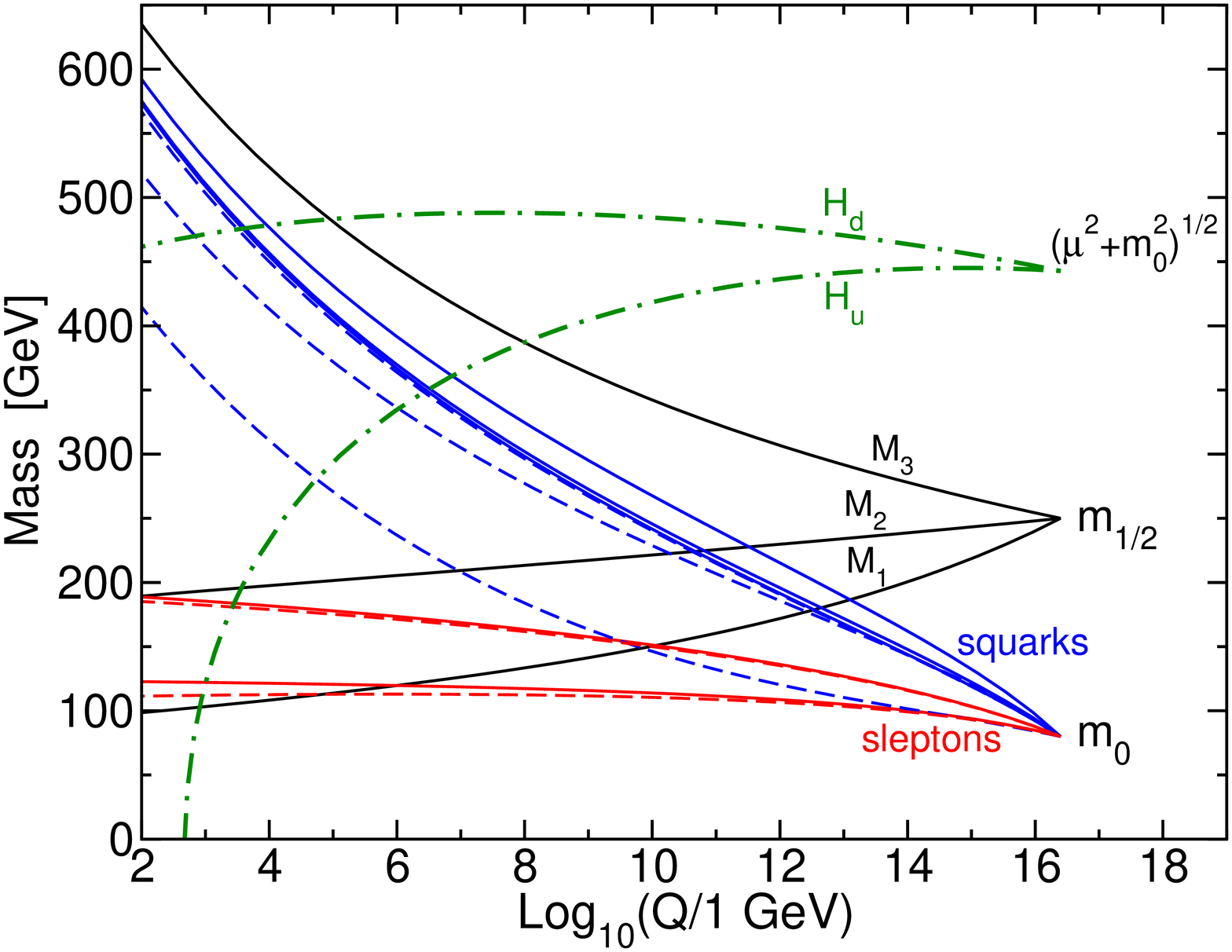}}
\end{minipage}
\vspace{5pt}
\caption[]{\small{(a) Left panel: Gauge coupling unification in MSSM
    \cite{uni}.  (b) Right panel: Negative $m_{H_u^2}$ in low energy triggers
    EWSB \cite{reviews}.}}
\end{figure*}

\subsubsection{Supersymmetry provides a cold dark matter candidate}

The present energy density (in units of critical density) of a thermal-relic
particle ($\chi$) is theoretically calculated as $\Omega_\chi h^2 \sim
0.1~{\rm pb}/\sigma$, where $\sigma$ is the thermal averaged non-relativistic
cross section for $\chi \chi \to f \bar{f}$. The observed dark matter density
is $\Omega_{\rm DM} h^2 = 0.114 \pm 0.003$ \cite{Hinshaw:2008kr}. Thus a
weakly interacting particle with a $\sim 100$ GeV mass, which has a
typical cross section of a pb, fits the bill! Interestingly, in the
theoretical formula, the coefficient 0.1 was derived using cosmological
parameters without any direct connection to the weak scale. This numerical
coincidence deserves attention. Supersymmetry with conserved $R$-parity
can provide such a dark matter candidate, a neutralino.

\subsubsection{Supersymmetry provides a framework to turn on gravity}

Local supersymmetry leads to supergravity. Thus gravity can be unified
with all other interactions. All string models invariably include
supersymmetry as an intergral part.

\subsection{The parameters in a general supersymmetric model}
In the minimal supersymmetric standard model (MSSM), where we do not assume
any particular mediation mechanism for its breaking and do not impose any GUT
conditions, the supersymmetry breaking soft parameters are not related to one
another. Here, we will see how a general supersymmetric model can be
parametrized \cite{feyn-rules}.

First consider the superpotential, written in terms of the `chiral
superfields', as
\begin{eqnarray}
W & = & \sum_{ij} \left(h_e^{ij} \L_i \H_d \E_j^c + h_d^{ij} \Q_i \H_d \D_j^c +
h_u^{ij} \Q_i \H_u \U_j^c\right) +
\mu \H_d \H_u.
\label{w}
\end{eqnarray}
Above, the sum is over the different generations.  $\H_d$ and $\H_u$ are the
two Higgs doublet superfields. The former gives masses to down-type quarks and
charged leptons and the latter gives masses to up-type quarks. $\L$ and $\Q$
are lepton and quark doublet superfields; $\E^c$, $\D^c$ and $\U^c$ are the
singlet charged lepton, down quark and up quark superfields,
respectively. $h_e$, $h_d$ and $h_u$ are the Yukawa couplings and $\mu$ is the
Higgs mixing parameter. Symbols with hats mean superfields and without hats
refer to the corresponding scalar fields.

The Lagrangian is given by
\begin{eqnarray}
-L = \sum_i \left|\frac{\partial {W}}{\partial {\phi_i}}\right|^2
+ \sum_{ij}\frac{\partial^2{W}}{\partial \phi_i \partial \phi_j}
\psi_i \psi_j
+
\frac{1}{2} \sum_\alpha |D_\alpha|^2 + \sum_{ij\alpha} \sqrt{2}
g_\alpha \psi_i (T^\alpha)^i_j \phi_j^* \lambda_\alpha,
\label{l}
\end{eqnarray}
where $\phi_i$ and $\psi_i$ the generic scalar and fermion fields
within the $i$th chiral multiplet, and $\lambda_\alpha$ represents the
gaugino which is a Majorana fermion in the vector multiplet with
$\alpha$ as the gauge group index.  The $D$ term is given by $D_\alpha
= -g_\alpha \phi_i (T^\alpha)^i_j \phi_j^*$.

The soft breaking terms are given by ($i,j$: generation indices,
$\alpha$: gauge group label)
\begin{eqnarray}
-L_{\rm soft} & = &
\sum_{ij} \widetilde{m}_{ij}^2 \phi_i^*\phi_j +
\sum_{ij} \left(A_e^{ij} L_i H_d E_j^* + A_d^{ij} Q_i H_d D_j^* +
A_u^{ij} Q_i H_u U_j^*\right) \nonumber \\
& + & m_{H_d}^2 |H_d|^2 + m_{H_u}^2 |H_u|^2 +
(B_\mu H_d H_u + {\rm h.c.}) +
{1\over 2}\left(\sum_\alpha \widetilde{M}_\alpha \lambda_\alpha
\lambda_\alpha + {\rm h.c.}\right).
\label{soft}
\end{eqnarray}

\subsection{Counting parameters}
Let us now count the total number of real and imaginary parameters in the MSSM
\cite{dimo_sutter}.  Each Yukawa matrix $h_f$ in Eq.~(\ref{w}) has 9 real and
9 imaginary parameters, and there are 3 such matrices.  Similarly, each $A_f$
matrix in Eq.~(\ref{soft}) has 9 real and 9 imaginary parameters, and again
there are 3 such matrices.  The scalar mass square $\widetilde{m}_{ij}^2$ can
be written for 5 representations: $Q, L, U^c, D^c, E^c$. For each such
represenation, the $(3 \times 3)$ hermitian mass square matrix has 6 real and
3 imaginary parameters. Finally, we have 3 gauge couplings (3 real), 3 gaugino
masses (3 real and 3 imaginary), $\mu$ and $B_\mu$ parameters (2 real and 2
imaginary), $(m_{H_u}^2, m_{H_d}^2)$ (2 real), and $\theta_{\rm QCD}$ (1
real).  Summing up, there are 95 real and 74 imaginary parameters. But not all
of them are physical. If we switch off the Yukawa couplings and the soft
parameters, i.e., keep {\em only} gauge interactions, there is a global
symmetry, given by
\begin{equation}
 G_{\rm global} = U(3)^5 \otimes U(1)_{\rm PQ} \otimes U(1)_R.
\label{G}
\end{equation}
The Peccei-Quinn (PQ) and $R$ symmetries are global U(1) symmetries, which
will not be discussed any further. $U(3)^5$ implies that a unitary rotation
among the 3 generations for each of the 5 representations leaves the physics
invariant. However, this unitary symmetry is broken. Once a symmetry is
broken, the number of parameters required to describe the symmetry
transformation can be removed. For example, when a U(1) symmetry is broken, we
can remove one phase. Note that a U(3) matrix has 3 real and 6 imaginary
parameters. So we can remove 15 real and 30 imaginary parameters from the
Yukawa matrices once $U(3)^5$ is broken. As we will see, the PQ and $R$
symmetries are also broken. So we can remove 2 more imaginary parameters. But
even when all the Yukawa couplings and soft parameters are turned on, there is
still a global symmetry, 
\begin{equation}
G'_{\rm global} = U(1)_B \otimes U(1)_L,
\end{equation}
where $B$ and $L$ are baryon and lepton numbers. Hence we can remove {\em not}
32 but {\em only} 30 imaginary parameters.  So we are left with 95 $-$ 15 = 80
real and 74 $-$ 30 = 44 imaginary, i.e., a total of 124 independent
parameters. The SM had only 18 parameters. So broken supersymmetry gifts us
106 more!  In the SM we had only one CP violating phase. Now we have 43 new
phases which are CP violating! If we break $R$-parity, defined by $R_p =
(-)^{3B+L+2S}$, where $S$ is the spin, then we will have 48 more complex
parameters \cite{rpar}. The reason for having to deal with so many parameters
is that although we know how to parametrize broken supersymmetric theories
very well, we really do not know how the symmetry is actually broken. So
supersymmetry is not just a model, it is rather a class of models, each
scenario differing from the others by the way the parameters are related among
themselves. Once we subscribe to any given supersymmetry breaking mechanism,
e.g. supergravity, anomaly mediation, gauge mediation, gaugino mediation, and
so on, the number of independent parameters gets drastically reduced.

\subsection{Tree level Higgs spectrum and radiative correction}
MSSM contains two complex Higgs doublets for three good reasons: (i) to avoid
massless charged degrees of freedom, (ii) to maintain analyticity of the
superpotential, and (iii) to keep the theory free from chiral anomaly, which
requires two Higgs doublets with opposite hypercharges.  Out of the 8 degrees
of freedom they contain, 3 are `swallowed' by $W^\pm$ and $Z$, and the
remaining five give rise to 5 physical Higgs bosons -- two charged ($H^\pm$)
and three neutral. Of the three neutral ones, one is CP odd ($A$) and two are
CP even ($H$ and $h$). Their tree level masses are given by
\begin{eqnarray}
m_A^2 &  = & m_{H_u}^2 + m_{H_d}^2 + 2|\mu|^2, ~~
m_{H^\pm}^2  =  m_A^2 + M_W^2, \nonumber \\
m_{h,H}^2 & = & {1\over 2}\left[m_A^2 + M_Z^2 \mp \sqrt{(m_A^2 + M_Z^2)^2 -
    4m_A^2 M_Z^2 \cos^2 2\beta} \right]. 
\end{eqnarray}
Above, $\tan\beta = v_u/v_d$, where $v_u$ and $v_d$ are the two vevs of $H_u$
and $H_d$.  It follows that $m_{H^\pm} \geq M_W$, $m_{H} \geq M_Z$ and
$m_{h} \leq M_Z$ at tree level. Since scalar quartic coupling in
supersymmetry arises from gauge interaction ($D$-term), it is not unexpected
that the lightest Higgs at tree level is lighter than the $Z$ boson.  But the
radiative correction to $m_{h}^2$ grows as the fourth power of the top mass
and hence is quite large \cite{radcorr}:
\begin{eqnarray} 
\label{radmh}
m_{h}^2 \simeq M_Z^2 \cos^2 2\beta + \frac{3G_Fm_t^4}{\sqrt{2}\pi^2} 
\left[{\ln\left(\frac{m_{\tilde{t}}^2}{m_t^2} \right)} +
\frac{A_t^2}{M_S^2} \left(1- \frac{1}{12} \frac{A_t^2}{M_S^2} \right)\right].  
\end{eqnarray} 
Assumming that the supersymmetry breaking parameters $M_S$ and $A_t$ are in
the TeV range, the radiative correction pushes the upper limit on $m_{h}$ to
about 135 GeV. This constitutes a clinching test of supersymmetry. If a light
neutral Higgs is not found at LHC approximately within this limit, MSSM with
two Higgs doublet would be strongly disfavoured.

\subsection{Naturalness}
Large cancellation between apparently {\em unrelated} quantities yielding a
small physical observable is a sign of {\em weak health} of the theory.  A
theory is less `natural' if it is more `fine-tuned'. Now, to the point
\cite{naturalness,Feng:1999zg}. From the scalar potential minimization, we
obtain
\begin{eqnarray}
\frac{1}{2} M_Z^2 = \frac{m_{H_d}^2 - m_{H_u}^2 \tan^2\beta}
{\tan^2\beta - 1} - \mu^2,
\label{nat}
\end{eqnarray}
where $m_{H_u}^2 = m_{H_d}^2 - \Delta m^2$, where $\Delta m^2$ is the
correction due to RG running from the GUT scale to the electroweak scale. The
RG running is heavily influenced by the top quark Yukawa coupling.  EWSB
occurs when $m_{H_u}^2$ turns negative by way of $\Delta m^2$ overtaking
$m_{H_d}^2$ such that a cancellation between the two terms on the RHS of
Eq.~(\ref{nat}) exactly reproduces the LEP-measured $M_Z$ on the LHS.  This
refers to a cancellation between terms of completely different origin: the
first term on the RHS of Eq.~(\ref{nat}) involves soft scalar masses
parametrizing supersymmetry breaking, while the second term, i.e. the $\mu$
term, is supersymmetry preserving and appears in the superpotential.  How much
cancellation between these completely uncorrelated quantities can we tolerate?
Of course, this is an aesthetic criterion.  Barbieri and Giudice in
\cite{naturalness} introduced a measure 
\begin{eqnarray}
\Delta_i \equiv
\left|\frac{\partial M_Z^2/M_Z^2}{\partial a_i/a_i}\right|,
\label{delta}
\end{eqnarray}
where $a_i$ are input parameters at high scale. $\Delta$ is a measure
of fine-tuning. An upper limit on $\Delta$ can be translated into an
{\em upper} limit on superparticle masses.
\begin{figure*}
\begin{minipage}[t]{0.42\textwidth}
\epsfxsize=8cm
\centering{\epsfbox
{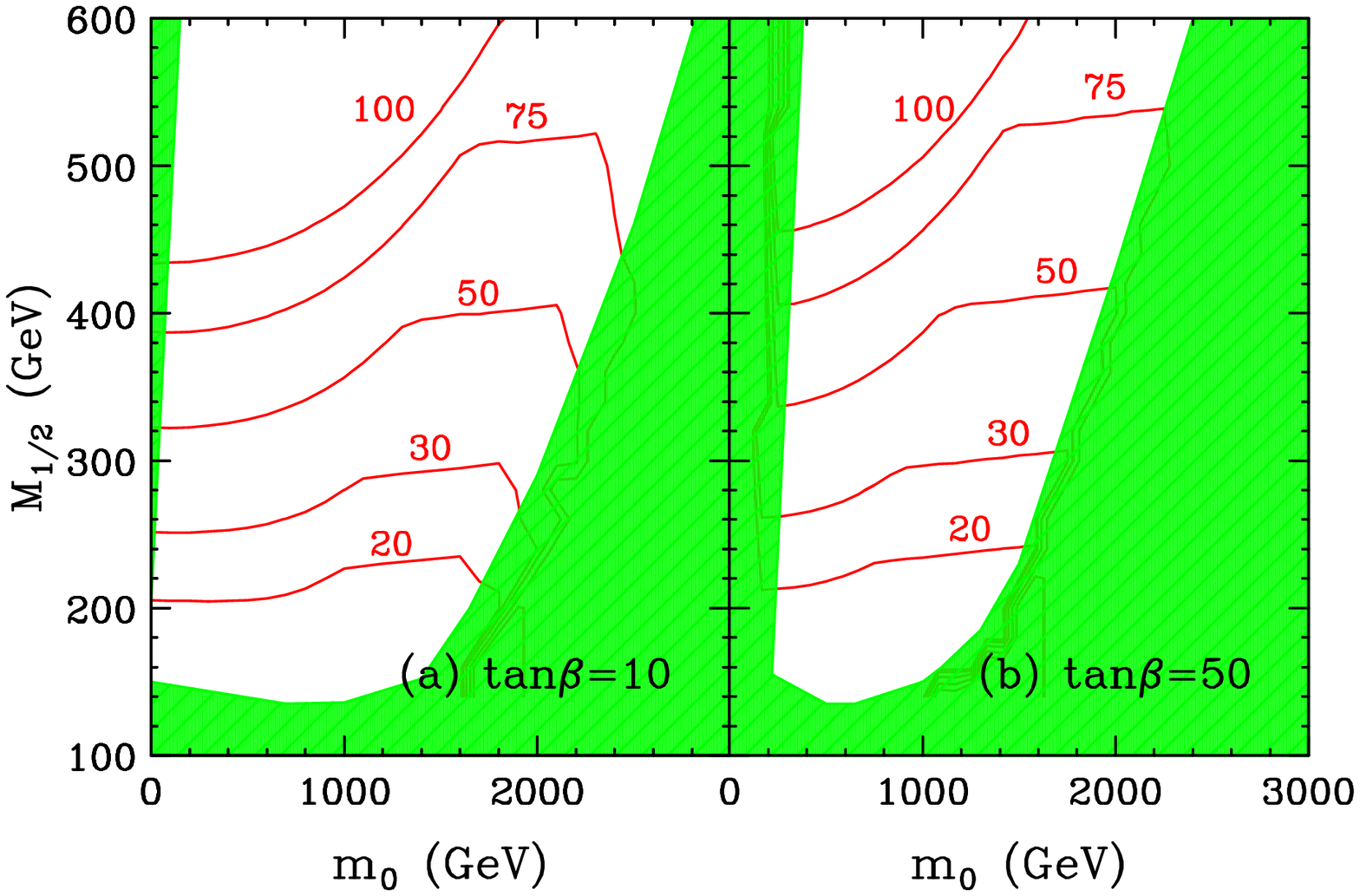}}
\end{minipage}
\hfill
\begin{minipage}[t]{0.42\textwidth}
\epsfxsize=8cm
\centering{\epsfbox
{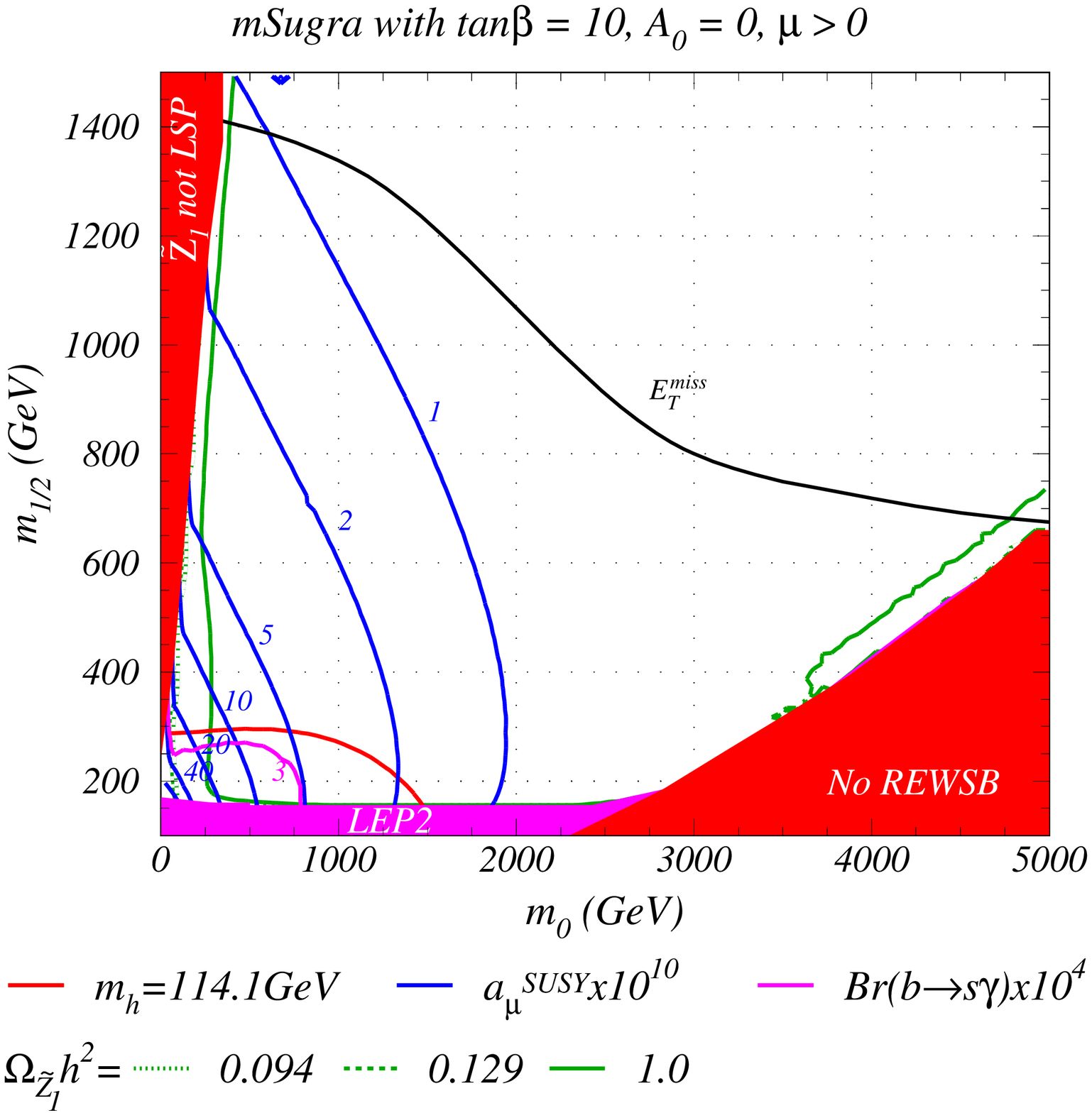}}
\end{minipage}
\vspace{5pt}
\caption[]{\small{(a) Left panel: Allowed region with different amount of
    fine-tuning \cite{Feng:1999zg}.  (b) Right panel: Only little is allowed
    \cite{Baer:2003wx} after including WMAP data.}}

\end{figure*}
Now we consider a specific example of fine-tuning in the context of
minimal supergravity. In this case, Eq.~(\ref{nat}) can be recast in the form  
\begin{equation}
M_Z^2 \simeq -2 |\mu^2| + \frac{3}{2\pi^2} h_t^2 m_{\tilde{t}}^2 
\ln\left(\frac{M_{\rm Pl}}{m_{\tilde{t}}}\right) 
= -2 |\mu^2| + {\cal{O}}(1) ~ m_{\tilde{t}}^2 .
\end{equation}
The `natural' expectation, i.e. without any fine-tuning, would be 
$M_Z \sim \mu \sim m_{\tilde{t}}$. On the other hand, the radiatively
corrected Higgs mass in Eq.~(\ref{radmh}) takes the approximate form 
\begin{equation}
m_h^2 \simeq M_Z^2 + \frac{3 h_t^2}{2 \pi^2} m_t^2 
\ln\left( \frac{m_{\tilde{t}}}{m_t}\right).
\end{equation}
Now, since $m_h > 114.4$ GeV, it automatically follows $m_{\tilde{t}} > 1$
TeV, contradicting the expectation from `naturalness', and implying a
fine-tuning or cancellation among unrelated parameters to the tune of a few
percent. This is called the `little hierarchy' problem of supersymmetry.  But
`little hierarchy' is after all a `little' hierarchy. At least, supersymmetry
solves the `gauge hierarchy problem' narrated earlier.


Figs.~4a and 4b refer to minimal supergravity. The left panel
\cite{Feng:1999zg} shows different contours for different values of the
fine-tuning parameter $\Delta$.  The contour for $\Delta=100$ admits a
cancellation as low as ($1/\Delta =$) 1\%, so it allows more parameter space
than the more conservative $\Delta=20$ curve which does not admit cancellation
below 5\%. The right panel shows that minimal supergravity is getting
increasingly squeezed as more data pour in. The WMAP data, in particular,
select out only a tiny region in the parameter space (for a detailed
description of this plot, see \cite{Baer:2003wx}). Relaxing certain
assumptions would definitely admit more parameter space.

\section{Little Higgs}

In Nature, we have seen light scalars before -- e.g. the pions -- though they
are composite. Their lightness owes to their pseudo-Goldstone nature. These
are Goldstone bosons which arise when the chiral symmetry group ${\rm SU(2)}_L
\times {\rm SU(2)}_R$ spontaneously breaks to the isospin group ${\rm
  SU(2)}_I$. A Goldstone scalar $\phi$ has a shift symmetry $\phi \to \phi +
c$, where $c$ is a constant, so any interaction which couples $\phi$ not as
$\partial_\mu \phi$ will break the Goldstone symmetry and attribute mass to
previously massless Goldstone. Quark masses and electromagnetic interaction
explicitly break the chiral symmetry. Electromagnetism attributes a mass to
$\pi^+$ of order $m_{\pi^+}^2 \sim (\alpha_{\rm em} /4\pi)\Lambda_{\rm
  QCD}^2$. Suppose, we conceive Higgs as a composite object, a
pseudo-Goldstone of some symmetry, and try to think of its mass generation in
the pion theme. We know that Yukawa interaction has a non-derivative Higgs
coupling, so it will break the Goldstone symmetry. Then, by analytic
continuation, if we replace $\alpha_{\rm em}$ by $\alpha_{\rm top}$ and
$\Lambda_{\rm QCD}$ by some $\Lambda_{\rm NP}$, we obtain $m_h^2 \sim
(\alpha_{\rm top} /4\pi)\Lambda_{\rm NP}^2$. The question is whether this
picture is phenomenologically acceptable. The answer is `no', as a $\sim$ 100
GeV Higgs would imply $\Lambda_{\rm NP} \sim 1$ TeV. Such a low cutoff is
strongly disfavoured by EWPT.

The little Higgs creators \cite{lh} had further tricks up their sleeves to
counter this obstacle. Consider the following all-order expansion in coupling
constants \cite{Rattazzi:2005di}:
\begin{equation} 
m_h^2  = \left(C_i \frac{\alpha_i}{4\pi}   
+ {C_i C_j\frac{\alpha_i \alpha_j}{(4\pi)^2}} + \cdot \cdot    
\right)~\Lambda_{\rm NP}^2  \, . 
\end{equation} 
Here, $\alpha_i$ are couplings of some external sources, e.g. gauge or Yukawa
interactions, that have nontrivial transformations under the Goldstone
symmetry. The coefficients $c_i$, $c_{ij}$ are symmetry factors. But now one
has to make such a smart choice of gauge groups and representation of scalars
that if any of the couplings ($\alpha_i$) vanishes the global symmetry is
partially restrored. Thus, to totally destroy the global symmetry one requires
the combined effect of at least two couplings.  Then the Goldstone acquires
mass parametrically at the 2-loop level:
\begin{equation}
m_h^2 \sim \left(\frac{\alpha}{4\pi}\right)^2 \Lambda_{\rm NP}^2 \, .    
\end{equation}
This is the concept of {\em collective symmetry breaking}.  Now, one can think
of new physics appearing at $\Lambda_{\rm NP} \sim$ 10 TeV scale. In a sense,
this is nothing but a {\em postponement} of the problem as the cutoff of the
theory is now 10 TeV instead of 1 TeV.
\begin{figure*}
\begin{minipage}[t]{0.42\textwidth}
\epsfxsize=5cm
\centering{\epsfbox
{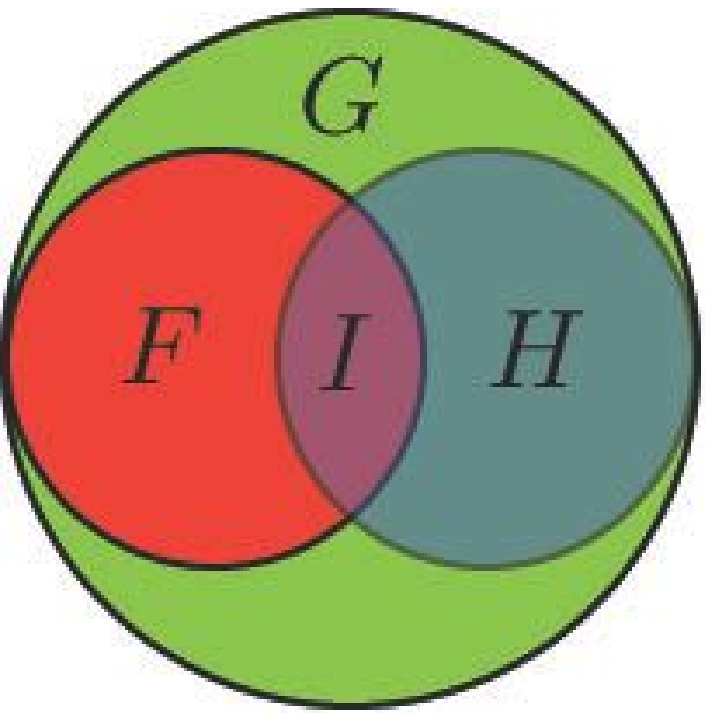}}
\end{minipage}
\hspace{0.5cm}
\begin{minipage}[t]{0.42\textwidth}
\epsfxsize=9cm
\centering{\epsfbox
{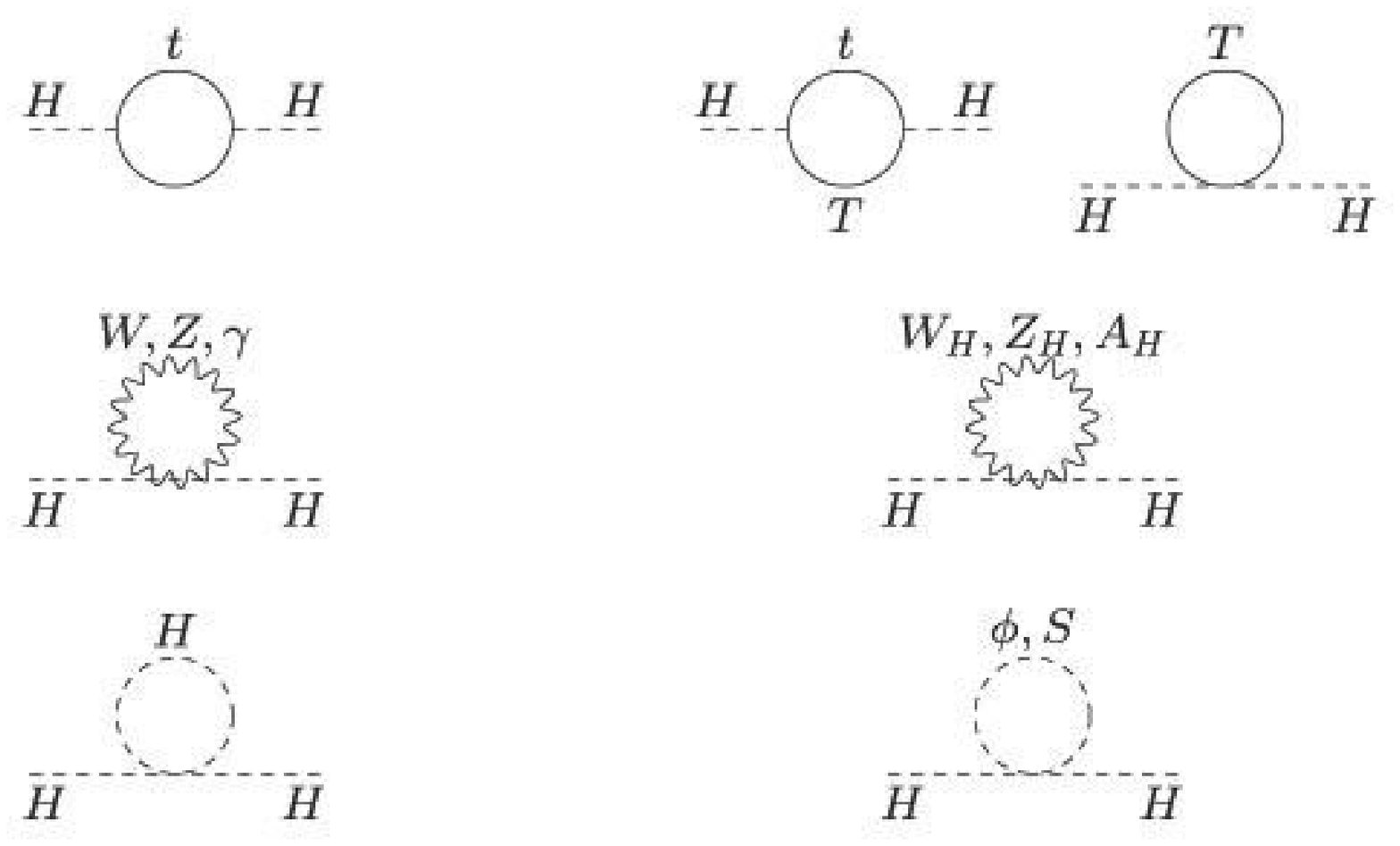}}
\end{minipage}
\vspace{5pt}
\caption[]{\small{(a) Left panel: A cartoon of little Higgs trick
    \cite{Cheng:2007bu}.  (b) Right panel: Cancellation between same
    statistics graphs \cite{Cheng:2007bu}.}}
\end{figure*}

To appreciate the little Higgs trick we look into Fig.~5 (left panel).
Consider a global group $G$ which spontaneously breaks to $H$ at a scale
$f$. The origin of this symmetry breaking is irrelevant below the cutoff scale
$\Lambda \sim 4\pi f$.  $H$ must contain SU(2) $\times$ U(1) as a subgroup so
that when a part of $G$, labelled $F$, is weakly gauged the unbroken SM group,
$I$, results.  The Higgs -- inside the doublet ($h$) under the SM group - is a
part of the Goldstone multiplet which parametrizes the coset space $G/H$. For
instance, G/H $=$ SU(5)/SO(5) scenario is called the `littlest', while G/H $=
{\rm SU(3)}^2 / {\rm SU(2)}^2$ scenario is called the `simplest'.  It is
important to note that the generators of the gauged part of $G$ do not commute
with the generators corresponding to the Higgs, and thus gauge (as well as
Yukawa) interactions break the Goldstone symmetry and induce Higgs mass at
one-loop level (parametrically at two-loop order, as we explained before).  A
clever construction of a little Higgs theory should have the following form of
the electroweak sector Higgs potential:
\begin{eqnarray}
V = -\mu^2 (h^\dagger h) + \lambda (h^\dagger h)^2, \nonumber
\end{eqnarray} 
where, the bilinear term is {\em suppressed}, $\mu^2 \sim \frac{g^4}{16\pi^2}
f^2 \ln (\Lambda^2/f^2)$, but, crucially, the quartic interaction should be
{\em unsuppressed}, $\lambda \sim g^2$.  

Since the gauge group is expanded, we have additional gauge bosons and
fermions. With the Higgs boson on external lines of a 2-point function, the
quadratic divergence arising from a $Z$-boson loop cancels against the same
from a $Z'$-boson loop; similar cancellation happens between a $t$-quark loop
and a $t'$-quark loop (see the right panel of Fig.~5). But the vev $f$
receives a quadratic correction, $f^2 \to F^2 = f^2 + (a/16\pi^2) \Lambda^2 =
(1+a) f^2$, where $a \sim {\cal{O}} (1)$. Thus, $m_h^2 = (g^4/16\pi^2) F^2 \ln
(\Lambda^2/F^2)$, where the quadratic sensitivity is shunned by a loop
suppression factor compared to the SM and this is where we gain
\cite{Kaul:2008cv}.  Clearly, $f \sim F \sim 1$ TeV. The cutoff of the theory
then becomes $\Lambda \sim 4\pi f$, which is 10 TeV (compared to 1 TeV in the
SM where naturalness breaks down). The `smoking gun' signals will constitute a
few weakly coupled particles (gauge bosons, top-like quark and a scalar
coupled to the Higgs) around $f \sim$ TeV.

\vspace{0.1cm}

It is worthwhile to compare and contrast supersymmetry and
little Higgs: 

\vspace{0.1cm}

\noindent \underline{Symmetry}: In supersymmetry, quadratic divergence to
Higgs mass-square cancels between loop diagrams containing different spin
particles. In little Higgs models, the above cancellation occurs between loop
diagrams with same spin particles. Although the quadratic sensitivity comes
back through $f$, it is accompanied by an extra suppression factor in the
Higgs mass, which implies that naturalness breaks down at $\sim$ 10 TeV. For
ultraviolet (UV) completion, one can either arrange successive little Higgs
mechanisms to push the naturalness scale further away, or, perhaps, appeal to
supersymmetry or technicolour to come in rescue at 10 TeV \cite{Kaul:2008cv}.

\vspace{0.1cm}

\noindent \underline{The minus sign}: In supersymmetry, large $h_t$ drives
$m_{H_u^2}$ negative triggering EWSB. In little Higgs models, too, large $h_t$
can generate the desired `negative sign'.

\vspace{0.1cm}

\noindent \underline{Fine-tuning}: $m_h > 114.4$ GeV requires large stop mass,
causing the `LEP paradox', leading to `little Hierarchy'. In the
next-to-minimal MSSM, there is an extra singlet which helps to ease this
tension. In little Higgs constructions, suppression of bilinear term compared
to quartic term requires fine-tuning in a large class of models.

\vspace{0.1cm}

\noindent \underline{Dark Matter}: The lightest supersymmetric particle
(e.g. the lightest neutralino) is an excellent dark matter candidate if
$R$-parity is exact.  In little Higgs models (the `littlest' type), one can
define a $T$-parity to distinguish between the SM particles ($T$-even) and the
extra species ($T$-odd). If $T$-parity is conserved, then the lightest $T$-odd
gauge boson is cosmologically stable and can act as a good dark matter
candidate.

\section{Gauge-Higgs unification}
The basic idea of Gauge-Higgs unification (GHU) is that the Higgs arises from
the internal components of a higher dimensional gauge field. Thus higher
dimensional gauge invariance provides a protection to the Higgs mass from
quadratic divergence.  When the extra coordinate is not simply connected
(e.g. $S^1$), there are Wilson line phases associated with the extra
dimensional component of the gauge field, conceptually similar to
Aharanov-Bohm phase in quantum mechanics. Their 4d fluctuation is identified
with the Higgs. There is no potential at the tree level, only through
radiative effects the Higgs boson acquires a mass.  The basic steps of
understanding the GHU mechanism are as follows:

\vspace{0.1cm}

\noindent ($i$)~ From a 4d point of view, a 5d gauge field $A_M$ can be
decomposed as $(A_\mu, A_5)$, where $\mu = 0,1,2,3$.  The idea is to relate
$A_5$ to the Higgs.  Consider a simple example: 5d QED on $S^1/Z_2$.  From a
4d point of view, the 5th component of the gauge field is indeed a scalar, and
there are $n$ such scalars, $A_5^{(n)}$. But none of these $A_5^{(n)}$
survives as a physical state. Each of them is `eaten up' by the corresponding
$A_\mu^{(n)}$, and the latter becomes massive.

\vspace{0.1cm}

\noindent ($ii$)~ Now take SU(3) as a gauge group and choose an orbifold
projection $P~=~{\rm diag}~(-1,-1,1)$ (in fundamental rep.)  which breaks
SU(3) to SU(2) $\times $ U(1).

Denote the SU(3) generators by $T^a$ where $a=1,...,8$.  Now, with a 
$Z_2$ projection, impose the conditions that the Lie-algebra valued $A_\mu
\equiv A_\mu^a T_a$ and $A_5 \equiv A_5^a T_a$ fields transform as $P
A_\mu P^\dagger = A_\mu$ and $P A_5 P^\dagger = -A_5$. 

Due to the relative {\em minus sign} between the two sets of transformations,
while the massless gauge bosons would transform in the adjoint of SU(2)
$\times $ U(1), the massless scalars would behave as a complex doublet under
SU(2) $\times $ U(1).  This complex doublet can be identified with our Higgs
doublet.

\vspace{0.1cm}

\noindent ($iii$)~ Indeed, the next question is how to generate the
scalar potential for electroweak breaking. The shift symmetry of
the scalar $A_5$ fields (i.e., the higher dimensional gauge invariance)
forbids us to write this potential at the tree level.

\vspace{0.1cm}

\noindent ($iv$)~ The interaction of the Higgs with bulk fermions and gauge
bosons will generate an effective scalar potential at one-loop level.  The
gauge loops tend to push $\langle A_5^0 \rangle$ to zero to minimize the
potential, while the fermionic loops tends to shift $\langle A_5^0 \rangle$
away from zero in the minimum of the potential. In fact, the KK fermions are
instrumental for generating the correct vev. This way of breaking SU(2)
$\times $ U(1) symmetry to U(1)$_{\rm em}$ is called the Hosotani mechanism
\cite{Hosotani:1983xw}. The one-loop Higgs mass is given by
\begin{eqnarray} 
m_h^2 \sim \frac{g^4}{128\pi^6} \frac{1}{R^2} \sum V''(\alpha),
\end{eqnarray}
where $\alpha$ is a dimensionless parameter arising from bulk interactions,
and the sum is over all KK particles. Clearly, 5d gauge symmetry is recovered
when $1/R \to 0$.

\vspace{0.1cm}

\noindent ($v$)~ A snapshot of the gauge spectrum is the following:
\begin{equation}
M_W^{(n)} = (n+\alpha)/R, ~~ M_Z^{(n)} = (n+2\alpha)/R, ~~
M_\gamma^{(n)} = n/R. 
\end{equation}

The periodicity property demands that the spectrum will remain
invariant under $\alpha \to \alpha+1$. This restricts $\alpha =
[0,1]$. Orbifolding further reduces it to  $\alpha =
[0,0.5]$. In principle, $\alpha$ can be fixed from the $W$ mass. 

\vspace{0.1cm}
{\em Admittedly, the above scenario does not phenomenologically work as it
  gives $M_Z^{(0)} = 2 M_W^{(0)}$. Yet, it provides an excellent illustration,
  providing clues to the right direction!}
\vspace{0.1cm}

We face some difficult obstacles while constructing a realistic scenario. The
GHU models often lead to ($a$) too small a top quark mass, ($b$) too small a
Higgs mass, and ($c$) too low a compactification scale. Besides, one has also
to worry about how to generate hierarchical Yukawa interaction starting from
higher dimensional gauge interaction which is after all universal. One way out
is to break the 5d Lorentz symmetry in the bulk:
\begin{eqnarray} 
L_g &=& -\frac{1}{4} F_{\mu\nu} F^{\mu\nu}  -\frac{\mathbf {a}}{4} F_{\mu 5}
F^{\mu 5}~~;~~ L_\Psi = \bar{\Psi} 
\left(i \gamma_\mu D^\mu - {\mathbf{k}} D_5 \gamma^5 \right)\Psi \, , 
\end{eqnarray} 
where the prefactors ${\mathbf {a}}$ and ${\mathbf{k}}$ need to be
phenomenologically tuned to match the data.  For detailed constructions of
gauge-Higgs unification scenarios, both in flat and warped space, we refer
the readers to Refs.~\cite{ghu}.

\section{Higgsless scenarios}
The idea is to trigger electroweak symmetry breaking without actually having a
physical Higgs. The mechanism relies on imposing different boundary conditions
(BC) on gauge fields in an extra-dimensional set-up.  The BCs can be carefully
chosen such that the rank of a gauge group can be lowered. The details can be
found in \cite{Csaki:2005vy,higgsless,radcorr_higgsless}. Here, we summarise
the essentials through the following steps:

\noindent ($i$)~ The simplest realisation is through the compactification of
the extra dimension on a circle with an orbifolding ($S^1/Z_2$). There are two
fixed points: $y = 0, \pi R$.

\vspace{0.1cm}

\noindent ($ii$)~ BC's: In general, $\partial_5 A_\mu^a = V A_\mu^a$, where
$V$ is the vev of a scalar field in a boundary. If $V = 0$ (called `Neumann
BC'), then $\partial_5 A_\mu^a = 0$. On the other hand, $V \to \infty$ (called
`Dirichlet BC') gives $A_\mu^a = 0$. So, some kind of a Higgs-like mechanism,
characterzied by $V$, is there in the backdrop, although eventually the gauge
bosons masses following EWSB induced by BC's would remain finite even in the
extreme limits of $V=0$ or $\infty$.

\vspace{0.1cm}

\noindent ($iii$)~ Appropriate BCs are chosen (for explicit formulae, see
\cite{higgsless}) which would ensure the
following gauge symmetry in bulk and in the two branes:\\
Bulk:~~
{${\rm SU(2)}_{\rm L} \times {\rm SU(2)}_{\rm R} 
\times {\rm U(1)}_{\rm B-L}$} \, , \\
$y = 0$ brane:~~ ${\rm SU(2)}_{\rm L} \times {\rm SU(2)}_{\rm R} \to
{\rm SU(2)}_{\rm D}$ \, , \\
$y = \pi R$ brane~~: ${\rm SU(2)}_{\rm R} \times {\rm U(1)}_{\rm B-L} \to {\rm
  U(1)}_{\rm Y}$ \, .

\vspace{0.1cm}

\noindent ($iv$)~ The gauge boson masses would follow from the solutions of
the transcendental equations involving $M_n$ and $R$. In flat space, the
following relations obtain: $M_\gamma = 0$, $M_W = (4R)^{-1}$, $M_Z = (\pi
R)^{-1} \tan^{-1}\sqrt{1+2g'^2/g^2}$.  But, $\Delta \rho \simeq 0.1$ is quite
large! The reason behind this large $\Delta \rho$ is that while the {\em
  custodial} symmetry is preserved in the bulk and at the $y=0$ brane, it is
violated at the $y=\pi R$ brane where the KK modes of the gauge fields have
sizable presence. One way out is to use the AdS/CFT correspondence in a warped
space.  Yet, in the above flat space example, relating the gauge boson masses
to the gauge couplings is quite an achievement and is a step to the right
direction.

\vspace{0.1cm}

\noindent ($v$)~ Recall that without a Higgs, unitarity violation would have
set in the SM at around a TeV. How do the Higgsless scenarios address this
issue? Here, the exchange of KK states retards the energy growth of the
$W_L$-$W_L$ scattering amplitude, {\em postponing} the violation of unitarity
in a {\em calculable} way beyond a TeV. Thus unitarity is kept partially under
control. This can be understood from a simple dimensional argument: In 4d, the
cutoff is $\Lambda_4 \sim 4\pi v \sim 1$ TeV. In 5d, the loop factor is
$g_5^2/24\pi^3 = g_4^2 R/12 \pi^2$, while the dimensionless quantity would be
$g_4^2 ER/12 \pi^2$. The 5d cutoff is then determined as $\Lambda_5 \sim
12\pi^2 /g_4^2 R \sim \Lambda_4 (3\pi/g_4^2) \sim 10$ TeV, for $1/R \sim v$.

\vspace{0.1cm}

\noindent ($vi$)~ EWPT poses a serious threat to the construction of a
realistic Higgsless model \cite{radcorr_higgsless}.

\section{Conclusions and Outlook}

\noindent {\bf 1.}~ All the BSM models we have considered are based on {\em
  calculability}. In all cases, the electroweak scale $M_Z$ can be expressed
in terms of some high scale parameters $a_i$, i.e. $M_Z = \Lambda_{\rm NP}
f(a_i)$, where $f(a_i)$ are calculable functions of physical parameters.

\vspace{0.1cm}

\noindent {\bf 2.}~ The new physics scales originate from different dynamics
in different cases: $\Lambda_{\rm SUSY} \sim M_S$ (the supersymmetry breaking
scale); $\Lambda_{\rm LH} \sim f \sim F$ (the vev of the pre-electroweak
Higgsing); $\Lambda_{\rm GHU} \sim R^{-1}$ (the inverse radius of
compactification).

\vspace{0.1cm}

\noindent {\bf 3.}~ In supersymmetry, the cutoff can be as high as the GUT or
the Planck scale. Both in little Higgs models and in the extra dimensional
scenarios (in general) the cutoff is much lower. In fact, in recent years
there is a revival of interest in strongly interacting light Higgs models
\cite{Giudice:2007fh}. However, their ultraviolet completion is an open
question!

\vspace{0.1cm}

\noindent {\bf 4.}~ In supersymmetry the cancellation of quadratic divergence
happens between a particle loop and a sparticle loop. Since a particle cannot
{\em mix} with a sparticle, the oblique electroweak corrections and the
$Zb\bar{b}$ vertex can be kept under control. In the relevant
non-supersymmetric scenarios, the cancellation occurs between loops with the
same spin states. Such states can {\em mix} among themselves, leading to
dangerous tree level contributions that EWPT either disapprove or, at best,
marginally admit.

\noindent {\bf 5.}~ Our goal is three-fold: (i) unitarize the theory, (ii)
sucessfully confront the EWPT, and (iii) maintain naturalness to the extent
possible. The tension arises as `naturalness' demands the spectrum to be
compressed, while `EWPT compatibility' pushes the {\em new} states away from
the SM states.

\vspace{0.1cm}

\noindent {\bf 6.}~ All said and done, the LHC is a `win-win' machine in terms
of discovery. Either we discover the Higgs, or, if it is not there, the new
resonances which would restore unitarity in gauge boson scattering are crying
out for verification. For the latter, we would need the super-LHC to cover the
entire parameter space.  In either scenario, we would need a linear collider
for precision studies.

\vspace{0.1cm}

\noindent{\bf Acknowledgements:}~ I thank the organizers for inviting me to
give this talk. I also thank Romesh Kaul for sharing his insights,
particularly in little Higgs models. I acknowledge hospitality at LPT-UMR
8627, Univ.~of Paris-Sud 11, Orsay, France, through a CNRS fellowship when
this review is being written up. This work is partially supported by the
project No.~2007/37/9/BRNS of BRNS (DAE), India.


\begin{thebibliography}{99}

\bibitem{rev_ewsb} G.~Altarelli, ``New Physics and the LHC,'' arXiv:0805.1992
  [hep-ph]; R.~Barbieri,
  ``Signatures of new physics at 14 TeV,''
  arXiv:0802.3988 [hep-ph]; 
G.~F.~Giudice,
  ``Theories for the Fermi Scale,''
  J.\ Phys.\ Conf.\ Ser.\  {\bf 110} (2008) 012014
  [arXiv:0710.3294 [hep-ph]]; 
R.~Barbieri, ``Searching for new physics at future accelerators,''
  Int.\ J.\ Mod.\ Phys.\ A {\bf 20} (2005) 5184 [arXiv:hep-ph/0410223]

\bibitem{Kaul:2008cv}
  R.~K.~Kaul,
  arXiv:0803.0381 [hep-ph].

\bibitem{Cheng:2007bu}
  H.~C.~Cheng,
  arXiv:0710.3407 [hep-ph].


\bibitem{Rattazzi:2005di}
  R.~Rattazzi,
  PoS {\bf HEP2005} (2006) 399
  [arXiv:hep-ph/0607058].


\bibitem{lepewwg} LEP Electroweak Working Group, {\sf
    http://lepewwg.web.cern.ch}.

\bibitem{Lee:1977eg}
  B.~W.~Lee, C.~Quigg and H.~B.~Thacker,
  Phys.\ Rev.\  D {\bf 16} (1977) 1519.

\bibitem{Altarelli:1994rb}
  G.~Altarelli and G.~Isidori,
  Phys.\ Lett.\  B {\bf 337} (1994) 141.

\bibitem{Kolda:2000wi}
  C.~F.~Kolda and H.~Murayama,
  JHEP {\bf 0007} (2000) 035
  [arXiv:hep-ph/0003170].


\bibitem{susy-books} See the text books on supersymmetry, \\
R.N. Mohapatra, `Unification and Supersymmetry: The Frontiers of
quark-lepton physics', Springer-Verlag, NY 1992;
M.~Drees, R.~Godbole and P.~Roy, `Theory and phenomenology of sparticles: An
account of four-dimensional N=1 supersymmetry in high energy physics,' World
Scientific (2004); H.~Baer and X.~Tata, `Weak scale supersymmetry: From
superfields to scattering events,' Cambridge, UK: Univ. Pr. (2006).


\bibitem{reviews}
For reviews on supersymmetry phenomenology, see for example,\\
S.P. Martin, 
hep-ph/9709356;
J.D. Lykken, 
hep-th/9612114;
P. Ramond, 
hep-th/9412234;
J. Bagger, 
hep-ph/9604232;
M. Drees, 
hep-ph/9611409;
S. Dawson, 
hep-ph/9612229;
J.F. Gunion, 
hep-ph/9704349;
X. Tata, 
hep-ph/9706307;
J. Louis, I. Brunner, S.J. Huber, 
hep-ph/9811341;
N. Polonsky, 
hep-ph/0108236;
I. Simonsen, 
hep-ph/9506369;
H.E. Haber, 
hep-ph/0103095;
A.~Djouadi,
  Phys.\ Rept.\  {\bf 459} (2008) 1
  [arXiv:hep-ph/0503173];
G. Bhattacharyya, 
hep-ph/0108267.

\bibitem{hall} For a discussion, see also, L. Hall, `The heavy top
  quark and supersymmetry', hep-ph/9605258.


\bibitem{intro-susy} E. Witten, Nucl. Phys. B 188 (1981) 513;
S. Dimopoulos, H. Georgi, Nucl. Phys. B 193 (1981) 150;
N. Sakai, Z. Phys. C 11 (1981) 153;
R.K. Kaul, Phys. Lett. 109 B (1982) 19;
R.K. Kaul, P. Majumdar, Nucl. Phys. B 199 (1982) 36.

\bibitem{uni}
J. Ellis, S. Kelley, D.V. Nanopoulos, Phys. Lett. B 260 (1991) 131;
U. Amaldi, W. de Boer, H. Furstenau, Phys. Lett. B 260 (1991)
447.

\bibitem{Hinshaw:2008kr}
  G.~Hinshaw {\it et al.}  [WMAP Collaboration],
  arXiv:0803.0732 [astro-ph].

\bibitem{feyn-rules}
H. Haber, G. Kane, Phys. Rept. 117 (1985) 75;
H.P. Nilles, Phys. Rept. 110 (1984) 1.

\bibitem{dimo_sutter} S. Dimopoulos, D.W. Sutter, Nucl. Phys. B 452
  (1995) 496.

\bibitem{rpar} R.~Barbier {\it et al.},
  Phys.\ Rept.\  {\bf 420} (2005) 1
  [arXiv:hep-ph/0406039]; G.~Bhattacharyya,
  arXiv:hep-ph/9709395; G.~Bhattacharyya,
  Nucl.\ Phys.\ Proc.\ Suppl.\  {\bf 52A} (1997) 83
  [arXiv:hep-ph/9608415].

\bibitem{radcorr}
J.~R.~Ellis, G.~Ridolfi and F.~Zwirner,
  Phys.\ Lett.\  B {\bf 257} (1991) 83 and Phys.\ Lett.\  B {\bf 262} (1991)
  477; Y.~Okada, M.~Yamaguchi and T.~Yanagida,
  Prog.\ Theor.\ Phys.\  {\bf 85} (1991) 1; H.~E.~Haber and R.~Hempfling,
  Phys.\ Rev.\ Lett.\  {\bf 66} (1991) 1815; 
A.~Brignole,
  Phys.\ Lett.\  B {\bf 281} (1992) 284;
M. S. Berger, Phys.\ Rev.\  {\bf D41} (1990) 225; 
J. F. Gunion and A. Turski, Phys.\ Rev.\ {\bf D39} (1989) 2701;
M.~Carena, J.~R.~Espinosa, M.~Quiros and C.~E.~M.~Wagner,
  Phys.\ Lett.\  B {\bf 355} (1995) 209
  [arXiv:hep-ph/9504316]; 
M.~Carena, M.~Quiros and C.~E.~M.~Wagner,
  Nucl.\ Phys.\  B {\bf 461} (1996) 407
  [arXiv:hep-ph/9508343]; 
H.~E.~Haber, R.~Hempfling and A.~H.~Hoang,
  Z.\ Phys.\  C {\bf 75} (1997) 539
  [arXiv:hep-ph/9609331];
S.~Heinemeyer, W.~Hollik and G.~Weiglein,
  Eur.\ Phys.\ J.\  C {\bf 9} (1999) 343
  [arXiv:hep-ph/9812472].



\bibitem{naturalness}
R. Barbieri, G.F. Giudice, Nucl. Phys. B 306 (1988) 63;
G. Anderson, G. Castano, Phys. Rev. D 52 (1995) 1693;
S. Dimopoulos, G.F. Giudice, Phys. Lett. B 357 (1995) 573;
G. Bhattacharyya, A. Romanino, Phys. Rev. D 55 (1997) 7015;
P. Ciafaloni, A. Strumia, Nucl. Phys. B 494 (1997) 41;
L. Giusti, A. Romanino, A. Strumia, Nucl. Phys. B 550 (1999) 3.
See also, G.~F.~Giudice,
  arXiv:0801.2562 [hep-ph].

\bibitem{Feng:1999zg}
  J.~L.~Feng, K.~T.~Matchev and T.~Moroi,
  Phys.\ Rev.\  D {\bf 61} (2000) 075005
  [arXiv:hep-ph/9909334].

\bibitem{Baer:2003wx}
  H.~Baer, C.~Balazs, A.~Belyaev, T.~Krupovnickas and X.~Tata,
  JHEP {\bf 0306} (2003) 054
  [arXiv:hep-ph/0304303].

\bibitem{lh}
N.~Arkani-Hamed, A.~G.~Cohen, E.~Katz and A.~E.~Nelson,
  JHEP {\bf 0207} (2002) 034
  [arXiv:hep-ph/0206021]; 
N.~Arkani-Hamed, A.~G.~Cohen, E.~Katz, A.~E.~Nelson, T.~Gregoire and 
J.~G.~Wacker,
  JHEP {\bf 0208} (2002) 021
  [arXiv:hep-ph/0206020]; 
N.~Arkani-Hamed, A.~G.~Cohen, T.~Gregoire and J.~G.~Wacker,
  JHEP {\bf 0208} (2002) 020
  [arXiv:hep-ph/0202089]; 
N.~Arkani-Hamed, A.~G.~Cohen and H.~Georgi,
  Phys.\ Lett.\  B {\bf 513} (2001) 232
  [arXiv:hep-ph/0105239];
D.~E.~Kaplan and M.~Schmaltz,
  JHEP {\bf 0310} (2003) 039
  [arXiv:hep-ph/0302049];
D.~E.~Kaplan, M.~Schmaltz and W.~Skiba,
  Phys.\ Rev.\  D {\bf 70} (2004) 075009
  [arXiv:hep-ph/0405257];
M.~Schmaltz,
  JHEP {\bf 0408} (2004) 056
  [arXiv:hep-ph/0407143];
M.~Schmaltz and D.~Tucker-Smith,
  Ann.\ Rev.\ Nucl.\ Part.\ Sci.\  {\bf 55} (2005) 229
  [arXiv:hep-ph/0502182]. 


\bibitem{Hosotani:1983xw}
  Y.~Hosotani,
  Phys.\ Lett.\  B {\bf 126} (1983) 309.

\bibitem{ghu} 
N.~Haba, S.~Matsumoto, N.~Okada and T.~Yamashita,
  arXiv:0802.3431 [hep-ph];
M.~Carena, A.~D.~Medina, B.~Panes, N.~R.~Shah and C.~E.~M.~Wagner,
  Phys.\ Rev.\  D {\bf 77} (2008) 076003
  [arXiv:0712.0095 [hep-ph]];
G.~Cacciapaglia, C.~Csaki and S.~C.~Park,
  JHEP {\bf 0603} (2006) 099
  [arXiv:hep-ph/0510366]; 
G.~Panico, M.~Serone and A.~Wulzer,
  Nucl.\ Phys.\  B {\bf 739} (2006) 186
  [arXiv:hep-ph/0510373].


\bibitem{Csaki:2005vy}
  For a review, see C.~Csaki, J.~Hubisz and P.~Meade,
  arXiv:hep-ph/0510275.

\bibitem{higgsless}
C.~Csaki, C.~Grojean, L.~Pilo and J.~Terning,
  Phys.\ Rev.\ Lett.\  {\bf 92} (2004) 101802
  [arXiv:hep-ph/0308038]; 
C.~Csaki, C.~Grojean, H.~Murayama, L.~Pilo and J.~Terning,
  Phys.\ Rev.\  D {\bf 69} (2004) 055006
  [arXiv:hep-ph/0305237];
C.~Csaki, C.~Grojean, J.~Hubisz, Y.~Shirman and J.~Terning,
  Phys.\ Rev.\  D {\bf 70} (2004) 015012
  [arXiv:hep-ph/0310355].


\bibitem{radcorr_higgsless}
  R.~Barbieri, A.~Pomarol and R.~Rattazzi,
  Phys.\ Lett.\  B {\bf 591} (2004) 141
  [arXiv:hep-ph/0310285];
G.~Cacciapaglia, C.~Csaki, C.~Grojean and J.~Terning,
  Phys.\ Rev.\  D {\bf 71} (2005) 035015
  [arXiv:hep-ph/0409126];
G.~Cacciapaglia, C.~Csaki, C.~Grojean and J.~Terning,
  Phys.\ Rev.\  D {\bf 70} (2004) 075014
  [arXiv:hep-ph/0401160].


\bibitem{Giudice:2007fh}
  G.~F.~Giudice, C.~Grojean, A.~Pomarol and R.~Rattazzi,
  JHEP {\bf 0706} (2007) 045
  [arXiv:hep-ph/0703164].

\end{thebibliography}
\end{document}